\documentclass[aps,pra,10pt,twocolumn,superscriptaddress,floatfix]{revtex4-2}
\usepackage{amsmath,amssymb,amsthm}
\usepackage{graphicx}
\usepackage{xcolor}
\usepackage{bm}
\usepackage{makecell}
\usepackage{hyperref}
\hypersetup{
  colorlinks=true,
  citecolor=blue,
  linkcolor=blue,
  urlcolor=blue,
  pdftitle={Fully-connected three-mode squeezed vacuum: Gaussian entanglement, steering, and collective photon subtraction}
}

\allowdisplaybreaks

\begin{document}

\title{Fully-connected three-mode squeezed vacuum: Gaussian entanglement, steering, and collective photon subtraction}

\author{Manjia Mai}
\affiliation{Center for Quantum Science and Technology, Jiangxi Normal University, Nanchang 330022, China}
\author{Jifeng Sun}
\affiliation{Center for Quantum Science and Technology, Jiangxi Normal University, Nanchang 330022, China}
\author{Teng Zhao}
\affiliation{Center for Quantum Science and Technology, Jiangxi Normal University, Nanchang 330022, China}
\author{Ming Zhang}
\email{mingzhang@jxnu.edu.cn}
\affiliation{Center for Quantum Science and Technology, Jiangxi Normal University, Nanchang 330022, China}
\author{Liyun Hu}
\email{hlyun2008@126.com}
\affiliation{Center for Quantum Science and Technology, Jiangxi Normal University, Nanchang 330022, China}

\begin{abstract}
We investigate a fully-connected three-mode squeezed vacuum (FC-C3MSV) state, where all three modes are pairwise coupled through nonlinear interactions in a triangle ($K_3$) topology. Using the integration-within-ordered-product technique, we derive the normal product form of the squeezing operator and obtain the covariance matrix directly from the Bogoliubov transformation. Under symmetric coupling, the physical state is genuinely tripartite entangled for any nonzero squeezing, while the three Armstrong-type witnesses provide a finite-window sufficient experimental test; in the chain-type C3MSV only one of these witnesses is violated. We find that, despite two-mode entanglement, the fully-connected topology admits \emph{no} two-mode Gaussian steering ($\mathcal{G}^{i\to j}=0$) between any pair of physical modes; the steering resource is instead collective one-mode-versus-two steering $\mathcal{G}^{i\to jk}$, which is $\theta$-independent and grows with $r$. We analyze independent vacuum losses and obtain critical transmittances for steering survival: under full symmetric loss at $r=0.5$, one-to-two collective steering disappears at $\eta\approx0.58$, whereas reverse two-to-one collective steering survives down to $\eta\approx0.502$ and the underlying two-mode entanglement persists for all $\eta>0$. Finally, we revisit photon subtraction using a normalized phase-space derivation. A photon subtraction on a single physical mode does not generate Wigner negativity on another single mode, consistent with the absence of two-mode steering. Wigner negativity can instead be generated when Bob subtracts from the collective mode $(b+c)/\sqrt{2}$, with a loss threshold $\eta_c\approx0.667$ at $r=0.5$. These results distinguish pairwise and collective nonclassical resources in the FC-C3MSV and clarify the operational role of the complete-graph topology.
\end{abstract}

\maketitle

\section{Introduction}

Quantum entanglement, since the seminal work of Einstein, Podolsky, and Rosen~\cite{Einstein1935}, has been generalized to multipartite systems~\cite{Greenberger1989,Greenberger2007,Mermin1990,Dur2000,Briegel2001} and observed experimentally~\cite{Bouwmeester1999,Pan2000}. Such multipartite entanglement serves as a fundamental resource for quantum networks~\cite{Wehner2018}, enabling quantum teleportation~\cite{Zeilinger2018} and distributed quantum information processing with quantum advantages~\cite{Buhrman2010,McCutcheon2016}.

Nonlinear optics provides promising experimental tools for realizing multipartite entanglement, particularly four-wave mixing (FWM)~\cite{Yuen1979,Bondurant1984,Kumar1984,Reid1985,Gerry2005}. FWM in hot atomic vapor has been demonstrated as a reliable source of squeezed light~\cite{McCormick2007,Guo2014,Gupta2016,Swaim2017}. Recent advances utilize multiple pump beams~\cite{Jia2017,Cai2020,Dong2022}, spatially structured pumps~\cite{Wang2017,Swaim2018}, or cascaded setups~\cite{Wang2016,Wang2017b,Cao2017,Cai2015} to produce multimode quantum correlations.

Zhang and Glasser~\cite{Zhang2022} introduced a coupled three-mode squeezed vacuum (C3MSV) with chain-type coupling ($a$-$b$, $b$-$c$) generated via dual-pump FWM. Zhan~\emph{et~al.}~\cite{Zhan2023} systematically analyzed its Gaussian steering properties and remote Wigner negativity generation, and recent work has further applied C3MSV steering ideas in relativistic settings~\cite{Mi2026}. The chain-type Hamiltonian $H_I = i\hbar(\chi_1^* a_1a_2 + \chi_2^* a_2a_3 - \text{H.c.})$ features mode $b$ as a central hub, leading to unidirectional steering $b\to a,c$ and no steering between modes $a$ and $c$~\cite{Zhan2023}. Differences between this sign convention and Eq.~\eqref{eq:H} below amount to phase conventions for the squeezing parameters and do not affect the phase-independent quantities compared here.

\begin{figure}[t]
\centering
\includegraphics[width=0.9\columnwidth]{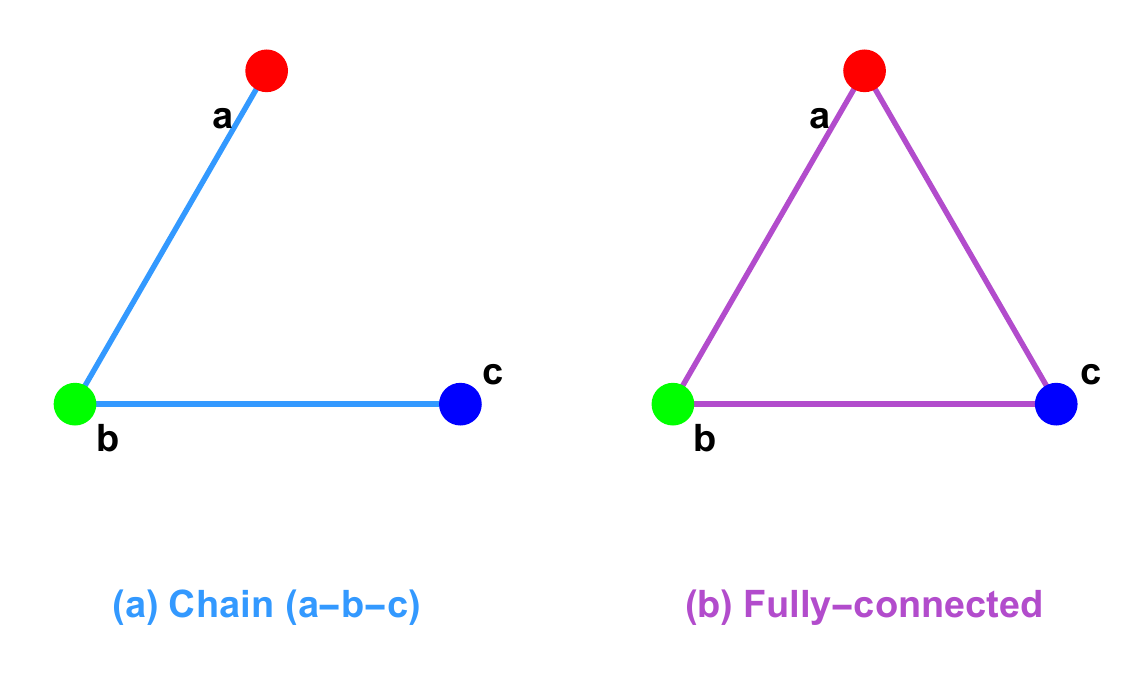}
\caption{(a) Chain-type coupling ($a$-$b$-$c$). (b) Fully-connected (triangle) coupling. Solid lines represent direct nonlinear interactions.}
\label{fig:topology}
\end{figure}

A more general configuration exists: fully-connected (FC) coupling where all three mode pairs interact ($a$-$b$, $b$-$c$, $a$-$c$), as illustrated in Fig.~\ref{fig:topology}. This triangle ($K_3$) topology could be approached experimentally with balanced multi-pump FWM geometries~\cite{Jia2017,Cai2020} and may offer distinctive features: full permutation symmetry, collective steering as a multipartite resource, and no central hub whose loss alone determines all steering directions. Teh~\emph{et~al.}~\cite{Teh2022,He2015} revealed the crucial role of network topology in steering distribution. The equal-coupling $K_3$ state provides an exactly solvable Gaussian example in which all physical two-mode reductions are entangled but none is Gaussian-steerable; the steerability is displaced into collective one-versus-two cuts. Operationally, this can be viewed as destructive interference, at the level of conditional covariance, between direct pair correlations and correlations mediated through the third mode. Fan and Jiang~\cite{Fan2002} studied three-mode squeezing in sum-frequency generation using the integration-within-ordered-product (IWOP) technique, but focused on a star-type structure, where one mode couples to the other two while the latter two do not couple directly. Although the fully symmetric squeeze matrix is contained in general multimode Gaussian-state theory and normal-ordering methods~\cite{Fan1999,Cariolaro2015}, its Gaussian-steering distribution, loss robustness, and collective photon-subtraction non-Gaussianity have not, to our knowledge, been systematically analyzed.

In this paper, we provide a complete theoretical analysis of the FC-C3MSV under symmetric coupling ($\chi_{12}=\chi_{13}=\chi_{23}$). Section~\ref{sec:basic} derives the Bogoliubov transformation, the covariance matrix (CM) from first principles, the normal product form via IWOP, and analyzes entanglement and steering. Section~\ref{sec:loss} studies decoherence under three loss configurations, obtaining the loss-degraded CM and analyzing the evolution of pairwise entanglement, one-to-two steering, and reverse two-to-one collective steering. Section~\ref{sec:wn} analyzes photon subtraction and shows that Wigner negativity is a collective-mode, rather than pairwise-mode, effect in the symmetric FC-C3MSV. Section~\ref{sec:conclusion} concludes. Technical derivations are provided in Appendices~\ref{app:A}--\ref{app:D}.

\section{Basic properties of the FC-C3MSV}
\label{sec:basic}

\subsection{Hamiltonian and Bogoliubov transformation}

The fully-connected interaction Hamiltonian is
\begin{equation}
\hat{H}_I = i\hbar\sum_{i<j}(\chi_{ij}\hat{a}_i^\dagger\hat{a}_j^\dagger - \chi_{ij}^*\hat{a}_i\hat{a}_j),\quad i,j=1,2,3,
\label{eq:H}
\end{equation}
with $[\hat{a}_i,\hat{a}_j^\dagger]=\delta_{ij}$. We focus on the symmetric case $\chi_{12}=\chi_{13}=\chi_{23}=\chi$. The dimensionless squeezing parameter is $\xi=\chi t_{\rm int}=re^{i\theta}$, where $t_{\rm int}$ is the interaction time, $r\ge0$ is the squeezing strength, and $\theta\in[0,2\pi)$ is the squeezing phase. The unitary squeezing operator is
\begin{equation}
S = \exp\!\Bigl[\xi\sum_{i<j}\hat{a}_i^\dagger\hat{a}_j^\dagger - \xi^*\sum_{i<j}\hat{a}_i\hat{a}_j\Bigr].
\label{eq:S}
\end{equation}

Introducing $\bm{a}=(a_1,a_2,a_3)^T$ and the $3\times3$ coupling matrix $M$ ($M_{ii}=0$, $M_{i\neq j}=\xi$), we have $S=\exp[\frac{1}{2}\bm{a}^\dagger M\bm{a}^{\dagger T} - \frac{1}{2}\bm{a}^T M^*\bm{a}]$.

The matrix $M$ is diagonalized by the orthogonal Fourier matrix $Q$ (Appendix~\ref{app:A}), with eigenvalues $2\xi$ (non-degenerate, eigenvector $\propto(1,1,1)^T$) and $-\xi$ (doubly degenerate). Following Fan and Zou's similarity transformation theory~\cite{Fan1999}, the Bogoliubov transformation $S^\dagger\bm{a}S = u\bm{a} + v\bm{a}^\dagger$ yields
\begin{align}
u &= \frac{1}{3}
\begin{pmatrix}
u_0 & u_1 & u_1 \\
u_1 & u_0 & u_1 \\
u_1 & u_1 & u_0
\end{pmatrix}, \label{eq:u}\\
v &= \frac{e^{i\theta}}{3}
\begin{pmatrix}
v_0 & v_1 & v_1 \\
v_1 & v_0 & v_1 \\
v_1 & v_1 & v_0
\end{pmatrix}. \label{eq:v}
\end{align}
with entries
\begin{align}
u_0 &= \cosh(2r) + 2\cosh r, \quad u_1 = \cosh(2r) - \cosh r, \label{eq:uv0}\\
v_0 &= \sinh(2r) - 2\sinh r, \quad\; v_1 = \sinh(2r) + \sinh r. \label{eq:uv1}
\end{align}
These satisfy the canonical Bogoliubov conditions $uu^\dagger-vv^\dagger=I$ and $uv^T=vu^T$~\cite{Cariolaro2015}. The mean photon number per mode is
\begin{equation}
\bar{n} = \langle a_i^\dagger a_i\rangle = \frac{v_0^2+2v_1^2}{9} = 2\sinh^2 r + \frac{4}{3}\sinh^4 r.
\label{eq:nbar}
\end{equation}
For $r\ll1$: $\bar{n}\approx 2r^2+O(r^4)$; for $r\gg1$: $\bar{n}\sim e^{4r}/12$.

\subsection{Normal product form via IWOP}

The normally ordered form of $S$ is derived via the IWOP technique~\cite{Fan1987,Fan1999,Fan2003} (detailed in Appendix~\ref{app:B}):
\begin{equation}
S = \frac{1}{\cosh r\,\sqrt{\cosh 2r}}\;:\!\exp\!\bigl( \mathcal{P} + \mathcal{N} - \mathcal{P}^\dagger \bigr)\!:,
\label{eq:normal_S}
\end{equation}
\begin{widetext}
where, defining $s_r\equiv\operatorname{sech}r$ and $s_{2r}\equiv\operatorname{sech}2r$ for brevity,
\begin{align}
\mathcal{P} &= \frac{e^{i\theta}(\tanh2r-2\tanh r)}{6}\sum_i\hat{a}_i^{\dagger2}
            + \frac{e^{i\theta}(\tanh r+\tanh2r)}{3}\sum_{i<j}\hat{a}_i^\dagger\hat{a}_j^\dagger, \\[2pt]
\mathcal{N} &= \frac{2s_r+s_{2r}-3}{3}\sum_i\hat{a}_i^\dagger\hat{a}_i
            + \frac{s_{2r}-s_r}{3}\sum_{i\neq j}\hat{a}_i^\dagger\hat{a}_j, \\[2pt]
\mathcal{P}^\dagger &= \frac{e^{-i\theta}(\tanh2r-2\tanh r)}{6}\sum_i\hat{a}_i^{2}
                     + \frac{e^{-i\theta}(\tanh r+\tanh2r)}{3}\sum_{i<j}\hat{a}_i\hat{a}_j.
\end{align}
\end{widetext}

\subsection{Covariance matrix}

Define quadratures $\hat{x}_j=(\hat{a}_j+\hat{a}_j^\dagger)/\sqrt{2}$, $\hat{p}_j=(\hat{a}_j-\hat{a}_j^\dagger)/(i\sqrt{2})$, and $\hat{\bm{R}}=(\hat{x}_1,\hat{p}_1,\hat{x}_2,\hat{p}_2,\hat{x}_3,\hat{p}_3)^T$. The CM $V_{ij}=\frac{1}{2}\langle\{\hat{R}_i,\hat{R}_j\}\rangle$ for the FC-C3MSV $|\psi\rangle=S|0\rangle$ is derived directly from the Bogoliubov transformation (Appendix~\ref{app:C}). For $\theta=0$, it takes the block form
\begin{equation}
V = \begin{pmatrix} A & B & B \\ B & A & B \\ B & B & A \end{pmatrix},
\label{eq:CM}
\end{equation}
with $A,B$ diagonal $2\times2$ matrices:
\begin{align}
A &= \frac{1}{6}\begin{pmatrix} e^{4r}+2e^{-2r} & 0 \\ 0 & e^{-4r}+2e^{2r} \end{pmatrix}, \label{eq:A}\\
B &= \frac{1}{6}\begin{pmatrix} e^{4r}-e^{-2r} & 0 \\ 0 & e^{-4r}-e^{2r} \end{pmatrix}. \label{eq:B}
\end{align}
For general $\theta\neq0$, the phase of the Bogoliubov matrix $v=e^{i\theta}v_{\theta=0}$ induces a symplectic rotation of the CM by angle $\theta/2$ in each $2\times2$ quadrature block:
\begin{align}
A(\theta) &= R(\theta/2)\,A(0)\,R(\theta/2)^T, \nonumber\\
B(\theta) &= R(\theta/2)\,B(0)\,R(\theta/2)^T,
\label{eq:AB_theta}
\end{align}
where $R(\phi)=\bigl(\begin{smallmatrix}\cos\phi & -\sin\phi \\ \sin\phi & \cos\phi\end{smallmatrix}\bigr)$ and $A(0),B(0)$ are the diagonal $\theta=0$ blocks of Eqs.~\eqref{eq:A}--\eqref{eq:B}. Explicitly,
\begin{align}
A(\theta) &= \begin{pmatrix} a_x\cos^2\tfrac{\theta}{2}+a_p\sin^2\tfrac{\theta}{2} & (a_x-a_p)\sin\tfrac{\theta}{2}\cos\tfrac{\theta}{2} \\[2pt] (a_x-a_p)\sin\tfrac{\theta}{2}\cos\tfrac{\theta}{2} & a_x\sin^2\tfrac{\theta}{2}+a_p\cos^2\tfrac{\theta}{2} \end{pmatrix}, \\[4pt]
B(\theta) &= \begin{pmatrix} b_x\cos^2\tfrac{\theta}{2}+b_p\sin^2\tfrac{\theta}{2} & (b_x-b_p)\sin\tfrac{\theta}{2}\cos\tfrac{\theta}{2} \\[2pt] (b_x-b_p)\sin\tfrac{\theta}{2}\cos\tfrac{\theta}{2} & b_x\sin^2\tfrac{\theta}{2}+b_p\cos^2\tfrac{\theta}{2} \end{pmatrix},
\end{align}
with $a_x,a_p,b_x,b_p$ the $\theta=0$ values. The rotation preserves all symplectic invariants: $\det A(\theta)=a_x a_p$ and $\det(A+2B)=\det(A-B)=1/4$ for all $\theta$, confirming that the state remains pure. Consequently, all conclusions drawn at $\theta=0$---$\bar{\nu}_{2|1}=1/2$ (hence $\mathcal{G}^{i\to j}=0$), the $\theta$-independence of $\mathcal{G}^{i\to jk}$, and the pairwise-versus-collective photon-subtraction distinction discussed below---hold for arbitrary $\theta$. For $r\to0$: $A\to\frac{1}{2}I_2$, $B\to0$. In what follows, we denote the $\theta=0$ diagonal entries by $a_x\equiv A_{11}(0)$, $a_p\equiv A_{22}(0)$, $b_x\equiv B_{11}(0)$, $b_p\equiv B_{22}(0)$.

Having established the CM, we now proceed to analyze the entanglement and steering properties that follow from it.

\subsection{Genuine tripartite entanglement}

The all-$r$ entanglement statement is most directly obtained from the physical one-versus-two bipartitions. The state is pure, as shown by the symplectic eigenvalues in Appendix~\ref{app:C}. For a pure tripartite Gaussian state, genuine tripartite entanglement follows if the state is entangled across each one-versus-two partition. By permutation symmetry, it is enough to examine the reduced CM of one physical mode, $A=\operatorname{diag}(a_x,a_p)$. Its determinant is
\begin{equation}
\det A=a_xa_p=\frac{5+4\cosh(6r)}{36}.
\label{eq:detA_single}
\end{equation}
Therefore the local symplectic eigenvalue is
\begin{equation}
\nu_i=\sqrt{\det A}>\frac{1}{2}\qquad (r>0),
\end{equation}
with equality only for the vacuum. Equivalently, the smallest partially transposed symplectic eigenvalue for the bipartition $i|jk$ is
\begin{equation}
\tilde{\nu}_{i|jk}^{(-)}
=\nu_i-\sqrt{\nu_i^2-\frac{1}{4}}<\frac{1}{2}\qquad (r>0).
\label{eq:pt_single_vs_pair}
\end{equation}
Thus every physical bipartition $1|23$, $2|13$, and $3|12$ is entangled for any $r>0$. Since the global state is pure, this proves genuine tripartite entanglement for all nonzero squeezing.

The Armstrong-type variance-product witnesses give an experimentally accessible sufficient test for the same multipartite resource over a finite window. Armstrong~\emph{et~al.}~\cite{Armstrong2015} provided a criterion based on variance inequalities for specific linear combinations of quadratures. For the fully symmetric FC-C3MSV, the natural choice is
\begin{align}
\hat{U}_1 &= 2\hat{x}_1-\sqrt{2}(\hat{x}_2+\hat{x}_3),\; \hat{V}_1 = 2\hat{p}_1+\sqrt{2}(\hat{p}_2+\hat{p}_3),
\end{align}
and cyclic permutations. The commutator vanishes, $[\hat{U}_j,\hat{V}_j]=0$, so a violation of the following bound cannot be attributed to an ordinary Heisenberg uncertainty constraint. Instead, it reflects nonlocal correlations tested against the biseparability condition. Following the variance-product witness used in Ref.~\cite{Armstrong2015}, and translating to our quadrature convention $[\hat{x},\hat{p}]=i$ and the above normalization of $\hat U_j,\hat V_j$, the biseparable bound becomes $\langle(\Delta\hat{U}_j)^2\rangle\langle(\Delta\hat{V}_j)^2\rangle\ge4$. Violation of this inequality is therefore a sufficient witness of genuine tripartite entanglement. Direct computation from the CM yields, for $\theta=0$:
\begin{align}
\langle(\Delta\hat{U}_j)^2\rangle &= \frac{1}{3}\bigl[(2-\sqrt{2})^2 e^{4r} + (2+\sqrt{2})^2 e^{-2r}\bigr],\\
\langle(\Delta\hat{V}_j)^2\rangle &= \frac{1}{3}\bigl[(2+\sqrt{2})^2 e^{-4r} + (2-\sqrt{2})^2 e^{2r}\bigr].
\end{align}
The window boundaries are the positive roots of
\begin{equation}
\begin{aligned}
&\frac{1}{9}\bigl[(2-\sqrt{2})^2 e^{4r} + (2+\sqrt{2})^2 e^{-2r}\bigr]\\
&\quad\times
\bigl[(2+\sqrt{2})^2 e^{-4r} + (2-\sqrt{2})^2 e^{2r}\bigr]=4,
\end{aligned}
\label{eq:armstrong_boundary}
\end{equation}
which give $r=0.2668$ and $r=0.9084$ numerically. \textbf{All three Armstrong-type witnesses are violated simultaneously over the open window $0.2668<r<0.9084$}, covering the experimentally most relevant range of squeezing. This finite-window violation is a sufficient certification, not the domain of entanglement itself. This contrasts with the chain-type C3MSV~\cite{Zhan2023}, where only the central-mode-related Armstrong witness is violated.

\subsection{Cyclic steering in the FC-C3MSV}

For a bipartite Gaussian state, the steering from $A$ to $B$ is quantified by the Kogias-Adesso measure~\cite{Kogias2015}. In our CM convention where $V=\frac{1}{2}\langle\{R,R\}\rangle$ (vacuum CM is $(1/2)I_2$), the steering condition is $\bar{\nu}_j^{B|A}<1/2$ and the steering measure is
\begin{equation}
\mathcal{G}^{A\to B} = \max\Bigl\{0, -\sum_{j:\bar{\nu}_j^{B|A}<1/2}\ln(2\bar{\nu}_j^{B|A})\Bigr\},
\label{eq:steer_def}
\end{equation}
where $\bar{\nu}_j^{B|A}$ are symplectic eigenvalues of $\sigma_{B|A}=V_B-V_{AB}^T V_A^{-1}V_{AB}$. (In the alternative convention $V=\langle\{R,R\}\rangle$, the condition is $\bar{\nu}<1$ and the measure is $-\ln\bar{\nu}$; the two are equivalent.)

\subsubsection{Two-mode steering}

For any pair of modes, the reduced two-mode CM is $\sigma_{12}=\begin{pmatrix}A&B\\B&A\end{pmatrix}$. The Schur complement $\sigma_{2|1}=A-BA^{-1}B$ is diagonal for $\theta=0$ with elements $a_x-b_x^2/a_x$ and $a_p-b_p^2/a_p$. Using Eqs.~\eqref{eq:A}--\eqref{eq:B}, a remarkable simplification occurs:
\begin{equation}
a_x - \frac{b_x^2}{a_x} = \frac{e^{-2r}(2e^{6r}+1)}{2(e^{6r}+2)},\qquad
a_p - \frac{b_p^2}{a_p} = \frac{e^{2r}(e^{6r}+2)}{2(2e^{6r}+1)}.
\end{equation}
Their product yields the symplectic eigenvalue
\begin{equation}
\boxed{\bar{\nu}_{2|1} = \sqrt{(a_x-b_x^2/a_x)(a_p-b_p^2/a_p)} = \frac{1}{2}},
\label{eq:nu_steer}
\end{equation}
which is \textbf{exactly $1/2$ for all $r>0$ and all $\theta$}. In our convention where the vacuum CM is $(1/2)I_2$, the Gaussian steering condition is $\bar{\nu}<1/2$ (the conditional variance must be \emph{smaller} than the vacuum variance for steering to exist). Since $\bar{\nu}_{2|1}=1/2$ saturates this bound, the Schur complement describes a minimum-uncertainty state: the measurement on mode 1 provides no reduction of the uncertainty of mode 2 below the vacuum level. Consequently, there is \textbf{no two-mode Gaussian steering} between any pair of modes in the FC-C3MSV:
\begin{equation}
\boxed{\mathcal{G}^{i\to j} = \mathcal{G}^{j\to i} = 0 \quad (\forall\, r>0,\;\forall\,\theta).}
\label{eq:G2}
\end{equation}
The same physical two-mode reduction is nevertheless entangled. For the reduced CM $\sigma_{12}=\left(\begin{smallmatrix}A&B\\B&A\end{smallmatrix}\right)$, the partially transposed symplectic eigenvalue is
\begin{align}
\nu_{\rm PPT}^{(-)}
&=\sqrt{\frac{\Delta_{\rm PT}-\sqrt{\Delta_{\rm PT}^2-4\mathcal{D}}}{2}},
\label{eq:ppt_pair}\\
\Delta_{\rm PT}&=2(a_xa_p-b_xb_p), \nonumber\\
\mathcal{D}&=(a_x^2-b_x^2)(a_p^2-b_p^2). \nonumber
\end{align}
Substitution of Eqs.~\eqref{eq:A}--\eqref{eq:B} gives $\nu_{\rm PPT}^{(-)}<1/2$ for every $r>0$; for example, $\nu_{\rm PPT}^{(-)}=0.3027$ at $r=0.5$. Thus the symmetric FC state realizes physical two-mode entanglement coexisting with zero two-mode Gaussian steering. The vanishing of two-mode steering is a joint consequence of the $K_3$ (complete graph) topology and the equal-coupling condition ($\chi_{12}=\chi_{13}=\chi_{23}$): under these combined symmetries the three-body correlations are so perfectly balanced that conditioning on one mode leaves the other in a minimum-uncertainty state. If the couplings are unequal (still $K_3$ but without full permutation symmetry), two-mode steering would generically be non-zero. The relation $\bar{\nu}_{2|1}=1/2$ also implies that the conditional state is a pure Gaussian state, a property that will be exploited in the Wigner negativity analysis of Sec.~\ref{sec:wn}.

\subsubsection{Physical origin of the vanishing two-mode steering}

The vanishing of two-mode steering can be understood in the Fourier basis. The symmetric mode $b_1$ (squeezed by $2r$) and the two antisymmetric modes $b_{2,3}$ (squeezed by $-r$) contribute to the reduced two-mode state with weights determined by $Q$. The Schur complement elements satisfy $(a_x-b_x^2/a_x)(a_p-b_p^2/a_p)=1/4$ identically, meaning the conditional state is always a minimum-uncertainty Gaussian state. The measurement on mode 1 provides exactly enough information to determine mode 2's state up to the Heisenberg limit, but no further. This ``pinning'' of the conditional variance at the vacuum level is a direct consequence of the $K_3$ symmetry: the three coupling channels contribute with weights that exactly balance, preventing any net reduction of uncertainty below $1/2$. In contrast, for a standard two-mode squeezed state (which lacks the third mode), the conditional variance decreases exponentially with $r$, yielding $\mathcal{G}\sim2r$.

\subsubsection{Collective steering}

For one mode versus the remaining two, the $4\times4$ Schur complement $\sigma_{23|1}$ has symplectic eigenvalues $\bar{\nu}_c$ (which depends on $r$) and $1/2$ (inherited from the degenerate antisymmetric subspace). The collective steering is
\begin{equation}
\boxed{\mathcal{G}^{i\to jk} = -\ln(2\bar{\nu}_c),}
\label{eq:Gcoll}
\end{equation}
which is independent of $\theta$ by symmetry and identical for all three modes. Here $\bar{\nu}_c<1/2$ is the non-trivial symplectic eigenvalue of $\sigma_{23|1}$; the second eigenvalue equals $1/2$ and contributes zero steering. The non-trivial eigenvalue can be written explicitly as $\bar{\nu}_c=3e^{3r}/[2\sqrt{e^{6r}+2}\sqrt{2e^{6r}+1}]$. Hence, for $r\ll1$, $\mathcal{G}^{i\to jk}=4r^2+O(r^4)$; for $r\gg1$, $\mathcal{G}^{i\to jk}=3r-\ln(3/\sqrt{2})+o(1)\sim3r$.

\begin{figure}[t]
\centering
\includegraphics[width=\columnwidth]{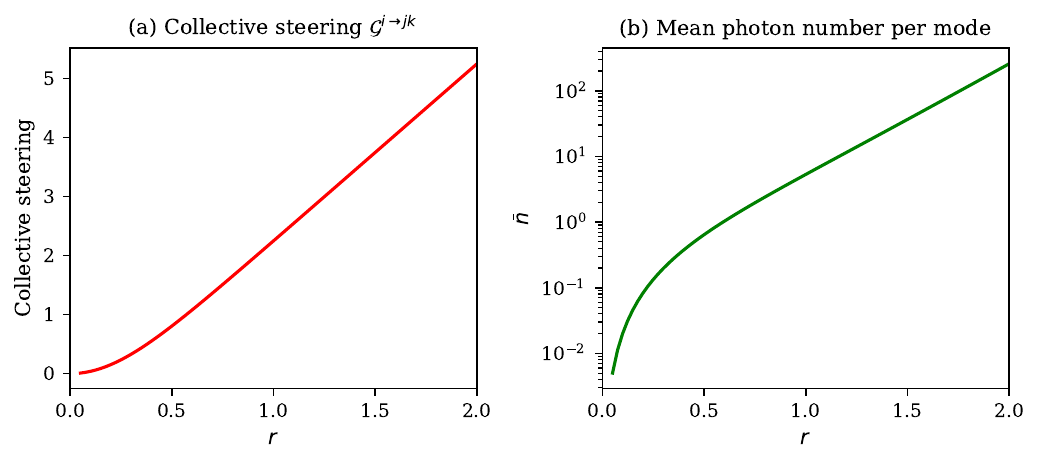}
\caption{(Color online) (a) Collective steering $\mathcal{G}^{i\to jk}$ as a function of $r$ at $\theta=0$. (b) Mean photon number per mode $\bar{n}=2\sinh^2 r+\frac{4}{3}\sinh^4 r$. The two-mode steering $\mathcal{G}^{i\to j}$ is identically zero for all $r$ and $\theta$.}
\label{fig:steering}
\end{figure}

Table~\ref{tab:steer} lists representative numerical values. The two-mode steering vanishes identically, while the collective steering $\mathcal{G}^{i\to jk}$ grows with $r$. For the lossless pure FC-C3MSV, the reverse collective steering $\mathcal{G}^{jk\to i}$ equals $\mathcal{G}^{i\to jk}$ across the same pure bipartition. Under loss, however, the state becomes mixed and the two directions generally differ; the reverse direction is therefore treated separately in Sec.~\ref{sec:loss}. Figure~\ref{fig:steering} shows the collective steering and the mean photon number as functions of $r$.

\subsubsection{Pairwise-versus-collective separation}

With $\mathcal{G}^{i\to j}=0$, the usual Gaussian-steering monogamy constraints~\cite{Reid2013} impose no non-trivial residual constraint on physical two-mode edges. The useful conclusion is instead structural: Gaussian steerability is not stored on any physical two-mode edge of the complete graph. It is delocalized across the network and appears only when one party can access a collective subsystem. In this sense the FC-C3MSV realizes a clean pairwise-versus-collective separation: pairwise entanglement is present, pairwise Gaussian steering is pinned exactly at the boundary, and the nonzero steering resource is collective. This is the network-steering role of the $K_3$ topology in the present continuous-variable setting.

\subsection{Comparison with chain-type C3MSV}

Tables~\ref{tab:compare} and~\ref{tab:quant} summarize the qualitative and quantitative differences between the two topologies at a representative squeezing parameter $r=0.5$, using the conventions specified in the table note. Both topologies exhibit vanishing two-mode steering for certain mode pairs (all pairs in FC, non-adjacent pairs in chain). The FC topology displays a different trade-off: it has no physical two-mode steering in the symmetric case, but it concentrates the steerability into a larger and permutation-symmetric collective resource at fixed parameter rather than fixed energy. Thus the larger FC collective-steering value at $r=0.5$ should not be read as a fixed-energy resource advantage.

\begin{table*}[t]
\caption{Comparison of chain-type~\cite{Zhan2023} and fully-connected C3MSV.}
\label{tab:compare}
\footnotesize\setlength{\tabcolsep}{3pt}
\begin{tabular*}{\textwidth}{@{\extracolsep{\fill}}|l|c|c|}
\hline
Property & Chain-type (1-2-3) & Fully-connected (FC-C3MSV) \\
\hline
Topology & Bus-type (linear) & Triangle (complete graph $K_3$) \\
Mode symmetry & Mode 2 central; 1,3 symmetric & All three fully symmetric \\
$\bar{n}$ per mode & $\bar{n}_2 > \bar{n}_1 = \bar{n}_3$ & $\bar{n}_1 = \bar{n}_2 = \bar{n}_3$ \\
Pairwise steering & $2\to1$, $2\to3$ unidirectional; $1\leftrightarrow3$ none & None: all vanish identically \\
Collective steering & Asymmetric & Symmetric, $\theta$-independent \\
Armstrong witness & 1 of 3 (mode 2-related) & All 3 simultaneously, finite window \\
CM off-diagonal & Two non-zero blocks ($B_{12},B_{23}$) & All three blocks equal ($B$) \\
Loss response & Sensitive to the central mode & Direction-dependent; collective resource \\
\hline
\end{tabular*}
\end{table*}

\begin{table*}[t]
\caption{Quantitative comparison at $r=0.5$, $\theta=0$. This fixed-parameter comparison is not energy-normalized.}
\label{tab:quant}
\centering
\footnotesize
\setlength{\tabcolsep}{4pt}
\begin{ruledtabular}
\begin{tabular*}{0.92\textwidth}{@{\extracolsep{\fill}}l|cc}
\hline
Quantity & Chain-type & FC-C3MSV \\
\hline
$\bar{n}_1=\bar{n}_3$ & $0.136$ & $0.641$ \\
$\bar{n}_2$ & $0.272$ & $0.641$ \\
$\mathcal{G}^{1\to2}$ & $0$ (none) & $0$ (none) \\
$\mathcal{G}^{2\to1}$ & $0.194$ (unidirectional) & $0$ (none) \\
$\mathcal{G}^{i\to jk}$ & $0.240$--$0.434$ & $0.808$ (all equal) \\
$\eta_{\text{crit}}$ (full sym. steering survival) & $0.50$--$0.667$ & $\approx0.58$ \\
$\eta_{\text{WN}}$ (Wigner-negativity survival) & not directly comparable & collective mode: $\approx0.667$ \\
\hline
\end{tabular*}
\end{ruledtabular}
\begin{minipage}{0.92\textwidth}
\footnotesize
For the chain-type state, the entries use the parameterization of Ref.~\cite{Zhan2023}, $r=\sqrt{r_1^2+r_2^2}$, with equal coupling $\phi=\pi/4$ so that $r_1=r_2=r/\sqrt{2}$. This $r$ is the effective squeezing strength used in Ref.~\cite{Zhan2023}, not the microscopic coupling rate in the Hamiltonian. The steering numbers are evaluated in the real-CM Kogias-Adesso convention used throughout this paper. In this convention, the center-to-edge steering is $\mathcal{G}^{2\to1}=-\ln[(c^2-s^2\cos2\phi)/(c^2+s^2)]=0.194$ at $r=0.5$. The chain-type Gaussian loss thresholds in the table are recalculated under the same full symmetric vacuum-loss model; Ref.~\cite{Zhan2023} discusses Wigner-negativity protocols in a different, lossless setting, so no chain-type Wigner-negativity loss threshold is quoted here.
\end{minipage}
\end{table*}

\begin{table}[t]
\caption{Steering measures for FC-C3MSV at $\theta=0$.}
\label{tab:steer}
\scriptsize
\setlength{\tabcolsep}{3.5pt}
\begin{ruledtabular}
\begin{tabular}{c|ccc}
$r$ & $\mathcal{G}^{i\to j}$ & $\mathcal{G}^{i\to jk}$ & $\bar{\nu}_{2|1}$ \\
\hline
0.2 & 0 & 0.154 & 0.500 \\
0.5 & 0 & 0.808 & 0.500 \\
0.8 & 0 & 1.658 & 0.500 \\
1.0 & 0 & 2.251 & 0.500 \\
1.5 & 0 & 3.748 & 0.500 \\
\end{tabular}
\end{ruledtabular}
\end{table}

The FC-C3MSV is fully symmetric under any mode permutation. It is genuinely tripartite entangled for all $r>0$, and the Armstrong-type criterion supplies an experimentally accessible finite-window witness in which all three inequalities are violated simultaneously. Despite physical two-mode entanglement, no two-mode Gaussian steering exists. Each mode can, however, steer the joint state of the other two with equal strength, independent of $\theta$. The collective steering is a direct consequence of the complete-graph topology. Since $\eta$ is a channel transmittance, a larger critical $\eta$ means a stricter transmission requirement, not a more permissive loss condition. Moreover, under the stated fixed-parameter conventions the FC state contains a larger mean photon number than the chain-type state. The comparison should therefore be interpreted as a topology comparison rather than an energy-normalized loss comparison.

\section{Entanglement and steering under vacuum loss}
\label{sec:loss}

We model losses as independent amplitude damping channels on each mode~\cite{Ferraro2005}, an idealization appropriate when the three spatial modes are well separated and interact with independent reservoirs. In this section and below, $\eta_j$ denotes the channel transmittance of mode $j$; when all lossy modes have the same transmittance we write the common value as $\eta$. This loss parameter is distinct from the coupling constants $\chi_{ij}$ and from the squeezing parameter $\xi=\chi t_{\rm int}$. In a realistic FWM setup with overlapping pump beams in a single atomic vapor cell, spatial mode correlations may introduce weakly correlated losses; the quantitative analysis of such correlations is deferred to future work. Under the independent-loss model, the CM evolves as
\begin{equation}
V_{\text{loss}} = T V T^T + N_{\text{loss}},
\qquad
N_{\text{loss}}=\bigoplus_{j=1}^3 \frac{1-\eta_j}{2}I_2,
\label{eq:Vloss}
\end{equation}
with $T=\bigoplus_{j=1}^3\sqrt{\eta_j}I_2$ and $\eta_j=e^{-\gamma_j \tau}$, where $\tau$ denotes the loss exposure time or propagation parameter. The block structure is preserved:
\begin{equation}
V_{\text{loss}} = \begin{pmatrix} A_1 & B_{12} & B_{13} \\ B_{12}^T & A_2 & B_{23} \\ B_{13}^T & B_{23}^T & A_3 \end{pmatrix},
\label{eq:Vloss_block}
\end{equation}
where $A_j = \eta_j A + \frac{1-\eta_j}{2}I_2$, $B_{jk} = \sqrt{\eta_j\eta_k}B$. We set $\theta=0$; the blocks $A,B$ are given by Eqs.~\eqref{eq:A}--\eqref{eq:B}.

\subsection{Three prototypical loss configurations}

We consider single-mode loss ($\eta_1=\eta$, $\eta_2=\eta_3=1$), two-mode loss ($\eta_1=\eta_2=\eta$, $\eta_3=1$), and full loss ($\eta_1=\eta_2=\eta_3=\eta$). For numerical illustration we take $r=0.5$, for which the lossless collective steering is $\mathcal{G}^{i\to jk}=0.808$. For this representative parameter, our numerical scan finds no physical two-mode Gaussian steering at $\theta=0$ in any of the three loss configurations. For a physical pair $i,j$, the conditional one-mode symplectic eigenvalue is
\begin{equation}
\bar{\nu}_{j|i}(\eta)=
\sqrt{\det\!\left[A_j-B_{ij}^T A_i^{-1}B_{ij}\right]}.
\label{eq:pairwise_loss_nu}
\end{equation}
Gaussian steering $i\to j$ exists only if $\bar{\nu}_{j|i}<1/2$. Under full symmetric loss the pairwise expression reduces to the explicit scalar check
\begin{equation}
\bar{\nu}_{\rm pair}^{\rm(fs)}(\eta,r)
=\sqrt{\left(a_x^{(\eta)}-\frac{[b_x^{(\eta)}]^2}{a_x^{(\eta)}}\right)
\left(a_p^{(\eta)}-\frac{[b_p^{(\eta)}]^2}{a_p^{(\eta)}}\right)},
\label{eq:pairwise_full_loss_nu}
\end{equation}
where $a_{x,p}^{(\eta)}=\eta a_{x,p}+(1-\eta)/2$ and $b_{x,p}^{(\eta)}=\eta b_{x,p}$. At the representative point $r=0.5$, Eq.~\eqref{eq:pairwise_full_loss_nu} gives $\bar{\nu}_{\rm pair}^{\rm(fs)}\in[0.500000,0.604092]$ over $\eta\in[10^{-6},1]$, so no physical two-mode Gaussian steering appears along the full-symmetric-loss path. Under single-mode and two-mode loss the block symmetry is broken, and the same no-steering conclusion is supported here only by the representative-parameter scan reported in Table~\ref{tab:pairwise_loss_check}. Table~\ref{tab:pairwise_loss_check} gives the reproducibility check over all six directed physical pairs. Consequently, for the $r=0.5$ loss analysis below, the physical two-mode steering remains zero in all three configurations considered; the nonzero steerability discussed below is a collective-mode resource.

\begin{table}[t]
\caption{Numerical check of pairwise Gaussian steering under loss, scanned only at the representative point $r=0.5$, $\theta=0$. The scan covers all six directed physical-mode pairs and $\eta\in[10^{-6},1]$ with 20001 uniform points.}
\label{tab:pairwise_loss_check}
\scriptsize
\setlength{\tabcolsep}{3pt}
\begin{ruledtabular}
\begin{tabular}{l|cc}
Loss case & Range of $\bar{\nu}_{j|i}$ & Steering \\
\hline
Single-mode loss & $[0.500000,\,1.121390]$ & 0 \\
Two-mode loss & $[0.500000,\,1.121390]$ & 0 \\
Full symmetric loss & $[0.500000,\,0.604092]$ & 0 \\
\hline
\end{tabular}
\end{ruledtabular}
\end{table}

For collective cuts, loss also separates the two Gaussian-steering directions. The one-to-two quantity $\mathcal{G}^{i\to jk}$ is obtained from the Schur complement of the two-mode block conditioned on mode $i$, as in Sec.~\ref{sec:basic}. The reverse direction, where the joint subsystem $(jk)$ steers the single mode $i$, is
\begin{align}
\mathcal{G}^{jk\to i}
&=\max\{0,-\ln(2\bar{\nu}_{i|jk})\},\\
\bar{\nu}_{i|jk}
&=\sqrt{\det\!\left[V_i-V_{i,jk}V_{jk}^{-1}V_{i,jk}^{T}\right]} .
\label{eq:reverse_collective_steering}
\end{align}
Here $V_i$ is the $2\times2$ single-mode block, $V_{jk}$ is the $4\times4$ two-mode block, and $V_{i,jk}$ is their $2\times4$ correlation block in the lossy CM.
For the lossless pure FC-C3MSV, $\mathcal{G}^{jk\to i}=\mathcal{G}^{i\to jk}$ by bipartite purity. Under full symmetric loss at $r=0.5$, however, the thresholds differ: the one-to-two direction dies at $\eta_{i\to jk}^{\rm crit}=0.5775\ldots$, whereas the reverse two-to-one direction survives down to $\eta_{jk\to i}^{\rm crit}=0.5022\ldots$. Thus there is a finite window, approximately $0.502<\eta<0.578$, in which the joint subsystem can still steer one mode although one-to-two steering has already vanished. Unless explicitly stated otherwise, the figures and tables below report the one-to-two quantity $\mathcal{G}^{i\to jk}$.

\begin{table*}[t]
\caption{One-to-two collective steering behavior under loss ($r=0.5$, $\theta=0$).}
\label{tab:loss}
\centering
\begin{ruledtabular}
\begin{tabular}{l|c|c|c}
\hline
 & Single-mode loss & Two-mode loss & Full loss \\
\hline
$\mathcal{G}^{1\to23}$ & Decays, dies at $\eta\approx0.50$ & Decays, dies at $\eta\approx0.50$ & Decays symmetrically \\
$\mathcal{G}^{3\to12}$ & Decays, survives $\forall\eta>0$ & Decays, survives $\forall\eta>0$ & Decays symmetrically \\
$\mathcal{G}^{2\to13}$ & Decays, survives $\forall\eta>0$ & Decays, dies at $\eta\approx0.50$ & Decays symmetrically \\
Survival threshold & Directional & Directional & $\eta_{i\to jk}^{\rm crit}\approx0.58$ (all one-to-two directions) \\
\hline
\end{tabular}
\end{ruledtabular}
\end{table*}

\textbf{Single-mode loss ($\eta_1=\eta$, $\eta_2=\eta_3=1$).} Mode 1 is lossy while modes 2,3 are lossless. The collective steering $\mathcal{G}^{1\to23}$ (lossy mode steering the other two) decays with $\eta$ and vanishes at $\eta\approx0.50$. The steering from the lossless modes, $\mathcal{G}^{2\to13}$ and $\mathcal{G}^{3\to12}$, decreases with $\eta$ but survives for all $\eta>0$, providing a directional asymmetry exploitable for communication protocols.

\textbf{Two-mode loss ($\eta_1=\eta_2=\eta$, $\eta_3=1$).} The steering $\mathcal{G}^{3\to12}$ (lossless mode 3) decreases with $\eta$ but survives, while $\mathcal{G}^{1\to23}$ and $\mathcal{G}^{2\to13}$ vanish at $\eta\approx0.50$. The single lossless mode preserves steerability in one direction only.

\textbf{Full loss ($\eta_1=\eta_2=\eta_3=\eta$).} All one-to-two collective steerings decay symmetrically, with a finite sudden-death threshold $\eta_{i\to jk}^{\rm crit}\approx0.58$ at $r=0.5$. Below this transmittance, no one-to-two collective steering survives; the reverse two-to-one direction remains nonzero down to $\eta_{jk\to i}^{\rm crit}\approx0.502$.

\begin{figure}[t]
\centering
\includegraphics[width=\columnwidth]{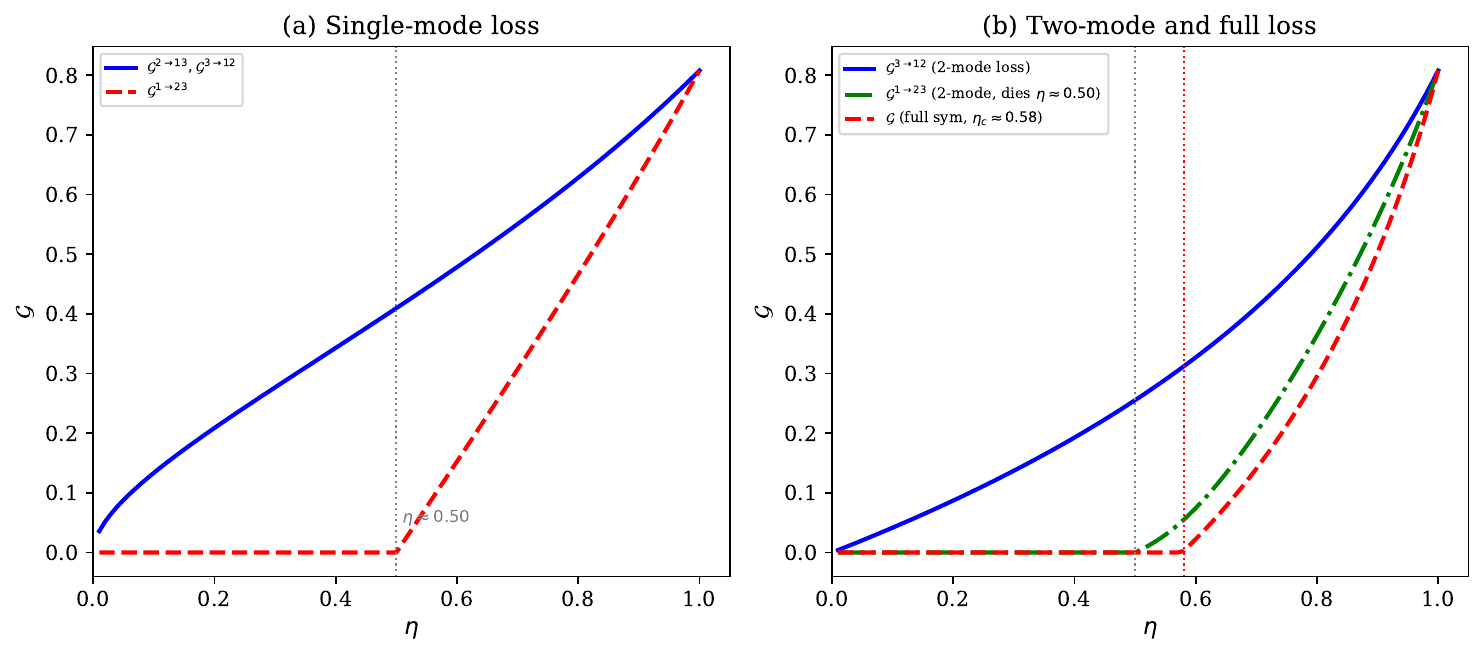}
\caption{(Color online) One-to-two collective steering under vacuum loss for $r=0.5$, $\theta=0$. (a) Single-mode loss on mode 1: $\mathcal{G}^{3\to12}$ and $\mathcal{G}^{2\to13}$ (blue solid, identical curves) from lossless modes survive for all $\eta$, while $\mathcal{G}^{1\to23}$ (red dashed, from lossy mode) vanishes at $\eta\approx0.50$. (b) Two-mode loss (blue solid: $\mathcal{G}^{3\to12}$ from lossless mode 3; green dash-dotted: $\mathcal{G}^{1\to23}$ and $\mathcal{G}^{2\to13}$, which die at $\eta\approx0.50$). Full symmetric loss (red dotted): all three one-to-two directions die at the common threshold $\eta\approx0.58$.}
\label{fig:loss}
\end{figure}

Figure~\ref{fig:loss} summarizes the one-to-two collective steering under the three loss configurations. In the single-mode loss case, $\mathcal{G}^{2\to13}$ and $\mathcal{G}^{3\to12}$ (steering from lossless modes) survive for all $\eta>0$, while $\mathcal{G}^{1\to23}$ (steering from the lossy mode) vanishes at $\eta\approx0.50$. In the two-mode loss case, only $\mathcal{G}^{3\to12}$ (steering from the single lossless mode) survives. In the full loss case, all three one-to-two directions vanish at the common threshold $\eta\approx0.58$.

\subsection{Entanglement versus steering under loss}

\begin{figure}[t]
\centering
\includegraphics[width=\columnwidth]{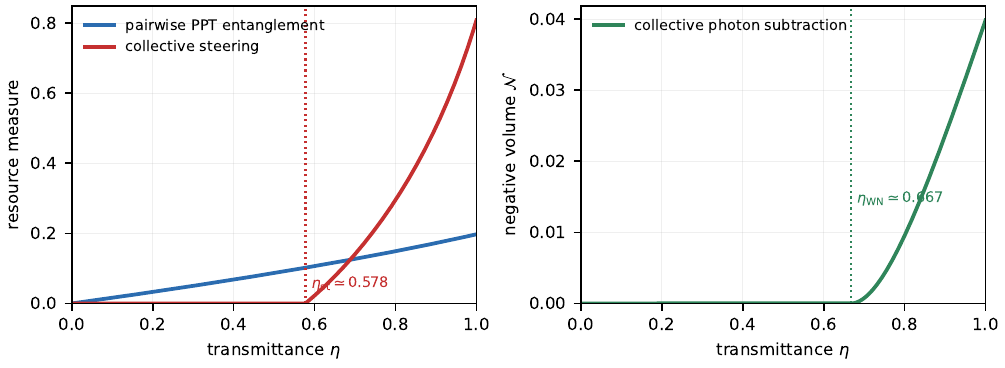}
\caption{(Color online) Loss robustness of pairwise and collective resources under full symmetric loss at $r=0.5$. (a) Physical pairwise PPT entanglement survives for all $\eta>0$, whereas the plotted one-to-two collective steering $\mathcal{G}^{i\to jk}$ dies at $\eta\approx0.58$. (b) The Wigner negativity generated by collective-mode photon subtraction vanishes at $\eta_{\rm WN}\approx0.667$.}
\label{fig:ent_steer}
\end{figure}

While the two-mode steering vanishes identically in the lossless case, the underlying two-mode entanglement---quantified by the PPT criterion $\nu_{\rm PPT}^{(-)}<1/2$---has a different loss threshold: for full symmetric loss at $r=0.5$, $\nu_{\rm PPT}^{(-)}$ remains below $1/2$ for all $\eta>0$, approaching the separability threshold only in the limit $\eta\to0$. This contrasts with one-to-two collective steering, which undergoes sudden death at $\eta\approx0.58$, the reverse two-to-one collective steering, which survives down to $\eta\approx0.502$, and the collective-mode Wigner negativity discussed in Sec.~\ref{sec:wn}, which vanishes at $\eta_{\rm WN}\approx0.667$. Figure~\ref{fig:ent_steer} compares the loss robustness of these distinct resources rather than a strict hierarchy within a single bipartition: the PPT curve refers to physical two-mode reductions, whereas the steering and Wigner-negativity curves involve collective one-versus-two or collective-mode operations. Along this symmetric loss path, Wigner negativity disappears first, one-to-two collective steering survives down to $\eta\approx0.58$, reverse collective steering survives further to $\eta\approx0.502$, and physical pairwise entanglement survives to arbitrarily low nonzero transmittance.

\subsection{Experimental relevance}

The loss response of the FC-C3MSV depends strongly on the loss configuration and the steering direction. For single-mode loss, the steering from the two lossless modes ($\mathcal{G}^{2\to13}\approx\mathcal{G}^{3\to12}$) survives for all $\eta>0$, while steering from the lossy mode vanishes at $\eta\approx0.50$. For two-mode loss, only $\mathcal{G}^{3\to12}$ survives for all $\eta>0$. For full symmetric loss at $r=0.5$, all three one-to-two directions share the common threshold $\eta\approx0.58$, whereas the reverse two-to-one collective direction survives to $\eta\approx0.502$. The one-to-two threshold corresponds to $\sim12$~km in low-loss optical fiber at telecom wavelengths if one uses $\eta=10^{-\alpha L/10}$ with $\alpha=0.2$~dB/km. Because $\eta$ denotes transmittance, this threshold should be read as the minimum transmission required for one-to-two collective steering survival under this model. Along the full symmetric loss path analyzed in Fig.~\ref{fig:ent_steer}, the physical two-mode PPT entanglement persists for all $\eta>0$, vanishing only in the limit $\eta\to0$; arbitrary asymmetric loss patterns are not claimed here.

A plausible implementation route is to use three overlapping pump beams in a single hot atomic vapor cell, building on recent multi-spatial-mode FWM experiments~\cite{Jia2017,Cai2020}. In this respect, Eq.~\eqref{eq:H} should be understood as an effective Gaussian-resource model: the three modes may correspond to spatial, frequency, polarization, or other experimentally separable degrees of freedom, provided the three pair-generation channels can be phase matched and phase locked. The symmetric coupling condition $\chi_{12}=\chi_{13}=\chi_{23}$ would require balancing the effective pump amplitudes, phases, and phase-matching conditions for the three pair-generation channels. Small symmetry breaking is expected to lift the exact pinning $\bar{\nu}_{2|1}=1/2$: physical two-mode steering may reappear and the collective steering will become direction-dependent. A quantitative tolerance analysis for coupling imbalance and phase mismatch requires a separate asymmetric-coupling model. At $r=0.5$, the diagonal normal-mode representation contains squeezing strengths $r$ and $2r$, corresponding ideally to about $4.3$~dB and $8.7$~dB in the relevant normal-mode quadratures, and the mean photon number per physical mode is $\bar{n}\approx0.64$. The full-symmetric-loss thresholds above should be read as total transmission requirements, including propagation, mode matching, and detection: at $r=0.5$, one-to-two collective steering requires an overall transmittance above about $0.58$, reverse two-to-one collective steering remains observable above about $0.502$, and collective-mode Wigner negativity requires an overall transmittance above about $0.667$. We note that verifying the one-to-two collective steering $\mathcal{G}^{i\to jk}$ requires joint homodyne measurements on two modes (e.g., modes $j$ and $k$), which can be implemented using two balanced homodyne detectors with a common local oscillator; experimentally this requires matched homodyne gains, stable relative LO phase, and calibrated shot-noise units in both detectors. The reverse quantity $\mathcal{G}^{jk\to i}$ uses the complementary conditional covariance of the single target mode and is experimentally reconstructed from the same three-mode covariance data. The collective-mode photon subtraction discussed below can be implemented by first combining modes $b$ and $c$ on a balanced beam splitter and then tapping the symmetric output with a weakly reflective beam splitter followed by a heralding single-photon detector. Imperfect visibility mixes the desired symmetric mode with the antisymmetric mode and increases the effective coefficient $C_d$ toward the non-negative physical-mode case, so the margin $C_d=-0.1996$ at $r=0.5$ should be regarded as the available negativity budget. The heralding probability depends on the tap reflectivity, detector efficiency, and the photon number in the tapped collective mode, and is therefore setup-dependent. At much larger squeezing the formal Gaussian-state predictions remain mathematically well defined, but practical FWM implementations may be limited by higher-order nonlinear processes and multipair emission; the moderate point $r=0.5$ is therefore the more realistic benchmark.

\section{Photon subtraction and Wigner negativity}
\label{sec:wn}

\subsection{Normalized phase-space expression}

The absence of two-mode steering in Sec.~\ref{sec:basic} has a direct consequence for remote non-Gaussian state preparation. Consider a two-mode zero-mean Gaussian state of the target mode $a$ and the measured mode $m$ with covariance matrix
\begin{equation}
V_{am}=
\begin{pmatrix}
 a_x&0&c_x&0\\
 0&a_p&0&c_p\\
 c_x&0&d_x&0\\
 0&c_p&0&d_p
\end{pmatrix}.
\label{eq:Vam}
\end{equation}
Photon subtraction on mode $m$ prepares the unnormalized conditional state
\begin{equation}
\tilde{\rho}_a=\operatorname{Tr}_m(\hat a_m\rho_{am}\hat a_m^\dagger)
              =\operatorname{Tr}_m(\hat n_m\rho_{am}).
\end{equation}
In the quadrature convention used throughout this work, the Weyl symbol of $\hat n_m$ is $(x_m^2+p_m^2-1)/2$~\cite{Kim2005}. Performing the Gaussian integral over the measured mode gives the normalized conditional Wigner function
\begin{equation}
W_a^{(m)}(x_a,p_a)=
\frac{W_a^G(x_a,p_a)}{P_m}
\left(C_m+K_xx_a^2+K_pp_a^2\right),
\label{eq:Wa_general}
\end{equation}
where
\begin{align}
W_a^G(x_a,p_a)
&=\frac{1}{2\pi\sqrt{a_xa_p}}
\exp\!\left[-\frac{x_a^2}{2a_x}-\frac{p_a^2}{2a_p}\right],\\
C_m&=\frac{1}{2}\left(d_{x|a}+d_{p|a}-1\right),\quad
P_m=\frac{d_x+d_p-1}{2},\label{eq:CmPm}\\
K_x&=\frac{1}{2}\left(\frac{c_x}{a_x}\right)^2,\quad
K_p=\frac{1}{2}\left(\frac{c_p}{a_p}\right)^2,\\
d_{x|a}&=d_x-\frac{c_x^2}{a_x},\quad
d_{p|a}=d_p-\frac{c_p^2}{a_p}.
\end{align}
Here $P_m=\langle \hat n_m\rangle$ is the ideal photon-subtraction normalization. The sign of $W_a^{(m)}$ at the phase-space origin is therefore determined only by $C_m$:
\begin{equation}
W_a^{(m)}(0,0)=
\frac{C_m}{2\pi\sqrt{a_xa_p}\,P_m}.
\label{eq:W0_general}
\end{equation}
The derivation of Eqs.~\eqref{eq:Wa_general}--\eqref{eq:W0_general}, including the negative-volume formula used below, is given in Appendix~\ref{app:D}. The normalized conditional state is defined for $P_m>0$, where $P_m$ is the mean photon number of the tapped mode. At the vacuum or complete-loss endpoint $P_m=0$, the ideal photon-subtraction event has zero probability and the conditional state is not a physically heralded normalized state.

\subsection{Why single physical-mode subtraction is insufficient}

For subtraction on one physical mode, say $m=b$, the two-mode covariance matrix has $d_x=a_x$, $d_p=a_p$, $c_x=b_x$, and $c_p=b_p$. Thus
\begin{equation}
C_b=\frac{1}{2}
\left(a_x-\frac{b_x^2}{a_x}
      +a_p-\frac{b_p^2}{a_p}-1\right).
\label{eq:Cb}
\end{equation}
The lossless FC-C3MSV gives
\begin{align}
d_{x|a}^{(b)}
&=\frac{e^{-2r}(2e^{6r}+1)}{2(e^{6r}+2)},\\
d_{p|a}^{(b)}
&=\frac{e^{2r}(e^{6r}+2)}{2(2e^{6r}+1)},
\end{align}
and hence $d_{x|a}^{(b)}d_{p|a}^{(b)}=1/4$. Therefore
\begin{equation}
d_{x|a}^{(b)}+d_{p|a}^{(b)}\ge 1,\qquad C_b\ge0,
\label{eq:Cb_positive}
\end{equation}
with equality only at $r=0$. At $r=0.5$ one finds
\begin{equation}
d_{x|a}^{(b)}=0.3429,\quad
d_{p|a}^{(b)}=0.7291,\quad
C_b=0.0360,
\end{equation}
so the conditional Wigner function is positive at the origin and, since $K_x,K_p\ge0$, non-negative everywhere. This result is not an accident of the lossless point. For a one-mode Gaussian conditional covariance, absence of Gaussian steering from $a$ to $b$ implies $\sqrt{\det V_{b|a}}\ge1/2$; for the diagonal form above, $\operatorname{Tr}V_{b|a}\ge2\sqrt{\det V_{b|a}}\ge1$, and therefore $C_b\ge0$. The absence of two-mode steering $\mathcal{G}^{a\to b}=0$ is thus consistent with the absence of Wigner negativity generated by subtracting a photon from a single physical mode~\cite{Walschaers2020,Walschaers2023}.

\subsection{Collective-mode photon subtraction}

The FC-C3MSV does possess collective steering from one mode to the other two. This suggests subtracting from Bob's symmetric collective mode
\begin{equation}
\hat d=\frac{\hat b+\hat c}{\sqrt{2}},
\end{equation}
which can be implemented by interfering modes $b$ and $c$ on a balanced beam splitter before the photon-subtraction tap. In the basis $(a,d)$, the covariance matrix has the same form as Eq.~\eqref{eq:Vam}, but with
\begin{equation}
d_x=a_x+b_x,\quad d_p=a_p+b_p,\quad
c_x=\sqrt{2}\,b_x,\quad c_p=\sqrt{2}\,b_p.
\label{eq:collective_entries}
\end{equation}
Consequently,
\begin{align}
C_d&=\frac{1}{2}\left(a_x+b_x-\frac{2b_x^2}{a_x}
      +a_p+b_p-\frac{2b_p^2}{a_p}-1\right),\label{eq:Cd}\\
P_d&=\frac{a_x+b_x+a_p+b_p-1}{2},\label{eq:Pd}\\
K_x&=\frac{b_x^2}{a_x^2},\qquad
K_p=\frac{b_p^2}{a_p^2}.
\end{align}
For $r=0.5$ and $\eta=1$,
\begin{equation}
\begin{aligned}
C_d&=-0.1996,\qquad P_d=1.0112,\\
W_a^{(d)}(0,0)&=-2.80\times10^{-2}.
\end{aligned}
\end{equation}
Thus the Wigner function has a genuine negative region around the origin. The key distinction is that the single physical mode $b$ does not receive enough conditional noise reduction from $a$, whereas the collective mode $d=(b+c)/\sqrt{2}$ does.

\subsection{Loss dependence and negative volume}

Under full symmetric loss,
\begin{equation}
a_{x,p}(\eta)=\eta a_{x,p}+\frac{1-\eta}{2},\qquad
b_{x,p}(\eta)=\eta b_{x,p}.
\end{equation}
The condition for a negative value at the origin is $C_d(\eta)<0$. Solving $C_d(\eta_c)=0$ at $r=0.5$ gives
\begin{equation}
\eta_c=0.6671605\ldots .
\label{eq:eta_c_wigner}
\end{equation}
For reference, Table~\ref{tab:loss_scaling} collects the full-symmetric-loss thresholds over representative squeezing values. The one-to-two thresholds decrease slowly with $r$, and throughout this range the ordering remains $\eta_{\rm WN}>\eta_{i\to jk}^{\rm crit}$. For the representative point $r=0.5$ used in Fig.~\ref{fig:ent_steer}, physical pairwise PPT entanglement survives for every nonzero transmittance along the symmetric loss path; this all-loss robustness is not universal at larger squeezing, so the table reports the pairwise PPT survival condition separately.

\begin{figure*}[t]
\centering
\includegraphics[width=0.72\textwidth]{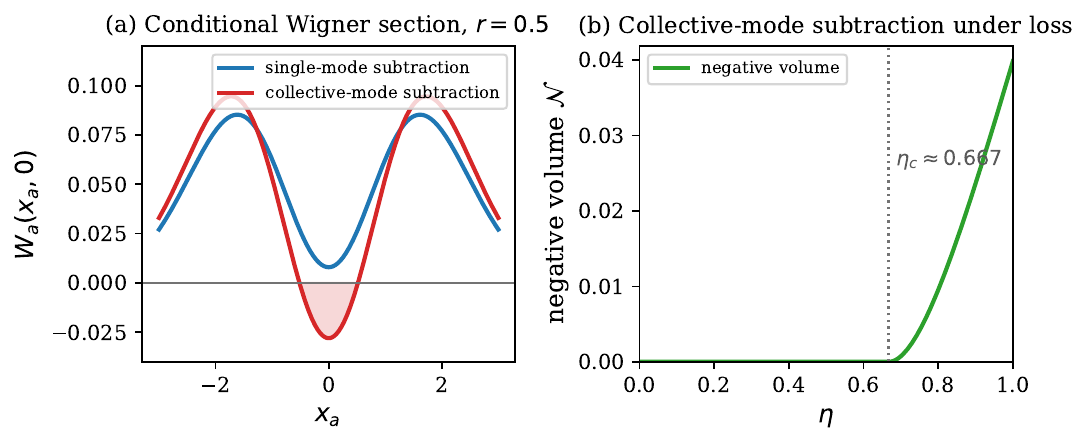}
\caption{(Color online) Photon subtraction on physical and collective modes at $r=0.5$. (a) Cross-section $W_a(x_a,0)$ in the lossless case. Subtraction on a physical mode $b$ leaves the Wigner function non-negative, whereas subtraction on the collective mode $d=(b+c)/\sqrt{2}$ produces a negative region around the origin. (b) Negative volume generated by collective-mode subtraction under full symmetric loss. The threshold is $\eta_c\simeq0.667$.}
\label{fig:wigner}
\end{figure*}

\begin{table}[!tbp]
\caption{Full-symmetric-loss thresholds versus squeezing. Here $\eta_{\rm crit}$ is the one-to-two collective Gaussian-steering threshold, $\eta_{\rm WN}$ is the collective-mode Wigner-negativity threshold, and the last column gives the physical pairwise PPT-entanglement survival condition.}
\label{tab:loss_scaling}
\scriptsize
\setlength{\tabcolsep}{4pt}
\begin{ruledtabular}
\begin{tabular}{c|ccc}
$r$ & $\eta_{\rm crit}$ & $\eta_{\rm WN}$ & pairwise PPT ent. survival \\
\hline
0.3 & 0.590 & 0.709 & all $\eta>0$ \\
0.5 & 0.578 & 0.667 & all $\eta>0$ \\
0.8 & 0.559 & 0.619 & all $\eta>0$ \\
1.0 & 0.550 & 0.598 & $\eta>0.479$ \\
\hline
\end{tabular}
\end{ruledtabular}
\end{table}
The numerical thresholds in Tables~\ref{tab:loss} and~\ref{tab:loss_scaling} were obtained from scalar root conditions: $\bar{\nu}_c(\eta,r)=1/2$ for one-to-two collective Gaussian steering, $\bar{\nu}_{i|jk}(\eta,r)=1/2$ for the reverse two-to-one check quoted in Sec.~\ref{sec:loss}, $C_d(\eta,r)=0$ for Wigner negativity, and Eq.~\eqref{eq:ppt_pair} with the full-symmetric-loss substitutions for pairwise PPT entanglement. We used bisection in $\eta$ with tolerance $10^{-8}$; the displayed thresholds are rounded to the quoted digits.
The negative volume is defined as~\cite{Kenfack2004}
\begin{equation}
\mathcal{N}=\int dx_a\,dp_a\,|W_a^{(d)}(x_a,p_a)|-1.
\end{equation}
For the anisotropic Gaussian polynomial in Eq.~\eqref{eq:Wa_general}, $\mathcal{N}$ is evaluated by the one-dimensional angular integral in Appendix~\ref{app:D}, using numerical quadrature over $\phi$; tightening the quadrature tolerance changes the tabulated values below the last quoted digit. Representative values are shown in Table~\ref{tab:wn}, and the corresponding Wigner-function behavior is plotted in Fig.~\ref{fig:wigner}.

\begin{table}[!tbp]
\caption{Collective-mode photon subtraction at $r=0.5$, $\theta=0$ under full symmetric loss.}
\label{tab:wn}
\scriptsize
\setlength{\tabcolsep}{4pt}
\begin{ruledtabular}
\begin{tabular}{c|ccc}
$\eta$ & $C_d$ & $W(0,0)$ & $\mathcal{N}$ \\
\hline
1.0 & $-0.1996$ & $-2.80\times10^{-2}$ & 0.0398 \\
0.9 & $-0.1336$ & $-2.20\times10^{-2}$ & 0.0236 \\
0.8 & $-0.0723$ & $-1.42\times10^{-2}$ & 0.0095 \\
$\eta_c$ & $0$ & $0$ & 0 \\
\hline
\end{tabular}
\end{ruledtabular}
\end{table}

\subsection{A simple remote-preparation protocol}

The collective-mode result above leads to a simple heralded protocol for remote preparation of Wigner negativity. Alice receives mode $a$, while Bob keeps modes $b$ and $c$ locally. Bob first applies a balanced beam splitter to form
\begin{equation}
\hat d=\frac{\hat b+\hat c}{\sqrt{2}},\qquad
\hat e=\frac{\hat b-\hat c}{\sqrt{2}}.
\label{eq:de_modes}
\end{equation}
For the symmetric FC-C3MSV, the covariance matrix in the mode ordering $(a,d,e)$ becomes
\begin{equation}
V_{ade}=
\begin{pmatrix}
A & \sqrt{2}B & 0\\
\sqrt{2}B & A+B & 0\\
0 & 0 & A-B
\end{pmatrix}.
\label{eq:Vade}
\end{equation}
Thus the antisymmetric mode $e$ is uncorrelated with the pair $(a,d)$ and can be ignored. All correlations relevant for remote Wigner-negativity generation are concentrated into the two-mode Gaussian state shared by Alice's mode $a$ and Bob's collective mode $d$.

Bob then taps the collective mode $d$ with a weakly reflective beam splitter of reflectivity $\kappa\ll1$ and monitors the reflected port with a single-photon detector. In the weak-tapping limit and neglecting dark counts, a click heralds the operation
\begin{equation}
\rho_{ad}\longmapsto
\frac{\hat d\rho_{ad}\hat d^\dagger}
{\operatorname{Tr}(\hat d^\dagger\hat d\rho_{ad})}.
\end{equation}
Bob communicates the heralding signal to Alice through a classical channel. Conditioned on this signal, Alice holds the remote non-Gaussian state
\begin{equation}
\rho_a^{\rm her}=
\frac{\operatorname{Tr}_d(\hat d\rho_{ad}\hat d^\dagger)}
{\operatorname{Tr}(\hat d^\dagger\hat d\rho_{ad})},
\label{eq:rho_a_her}
\end{equation}
whose Wigner function is precisely Eq.~\eqref{eq:Wa_general} with $m=d$. Therefore the state is Wigner negative whenever $C_d<0$. At $r=0.5$ under full symmetric loss, this requires $\eta>0.6671605\ldots$.

The click probability is setup-dependent. For weak tapping and detector efficiency $\zeta_{\rm det}$, it scales as
\begin{equation}
p_{\rm click}\simeq \kappa\,\zeta_{\rm det}\,P_d,
\end{equation}
up to corrections from multiphoton subtraction and dark counts, where $P_d=\langle \hat d^\dagger\hat d\rangle$ is given in Eq.~\eqref{eq:Pd}. At $r=0.5$ and unit transmission, $P_d\simeq1.01$; for a typical weak tap $\kappa\sim1\%$--$5\%$ and realistic detector efficiencies, this corresponds to a raw heralding probability in the percent-to-subpercent range before additional mode-matching and filtering losses. The protocol is experimentally modest in the following sense: it requires only a phase-stable balanced beam splitter on Bob's two modes, a standard weak-tap photon-subtraction module on one output port, classical heralding, and homodyne tomography on Alice's mode. Its main technical requirements are high interference visibility between $b$ and $c$, stable control of the relative phase selecting the symmetric mode $d$, low dark counts, and sufficiently small tapping reflectivity to suppress higher-order subtraction events. If the two Bob modes are not perfectly symmetric, the same idea can be generalized by subtracting from an optimized mode $\hat d_{\phi,\varphi}=\cos\phi\,\hat b+e^{i\varphi}\sin\phi\,\hat c$, with $\phi$ and $\varphi$ chosen to minimize the corresponding coefficient $C_m$ in Eq.~\eqref{eq:CmPm}.

Compared with the chain-type C3MSV of Ref.~\cite{Zhan2023}, the FC-C3MSV should therefore not be described as producing pairwise remote Wigner negativity by subtracting from a single physical output mode. Its non-Gaussian feature is collective: the complete graph creates a symmetric mode on Bob's side that is strongly correlated with Alice's mode. The comparison at fixed $r$ is also not energy-normalized, since the FC topology contains a larger mean photon number per physical mode. The qualitative conclusion is that, in the symmetric FC-C3MSV, both Gaussian steering and photon-subtraction-induced Wigner negativity are collective resources.

\section{Conclusion and outlook}
\label{sec:conclusion}

We have systematically investigated the FC-C3MSV, deriving its covariance matrix directly from the Bogoliubov transformation and using it to characterize the separation of resources created by the complete-graph topology. The main findings are: (i) the pure FC-C3MSV is genuinely tripartite entangled for every $r>0$, as follows from the entanglement of all physical one-versus-two bipartitions, while the Armstrong-type witnesses give a finite-window experimental test; (ii) despite physical two-mode entanglement, all two-mode Gaussian steering measures $\mathcal{G}^{i\to j}$ vanish in the symmetric FC state; (iii) the nonzero steering resource is collective: in the pure state $\mathcal{G}^{i\to jk}=\mathcal{G}^{jk\to i}$, while under loss the one-to-two and reverse two-to-one directions become inequivalent; (iv) independent vacuum losses produce direction-dependent steering thresholds, and along the full symmetric loss path at $r=0.5$ the one-to-two threshold is $\eta\approx0.58$, the reverse two-to-one threshold is $\eta\approx0.502$, and physical pairwise PPT entanglement persists for all nonzero transmittance; and (v) photon subtraction reflects the same pairwise-versus-collective distinction. A normalized Wigner-function derivation shows that subtracting from a single physical mode does not generate Wigner negativity on another physical mode, whereas subtracting from Bob's collective mode $(b+c)/\sqrt{2}$ does, with $\eta_c\approx0.667$ at $r=0.5$ under full symmetric loss.

These results identify the FC-C3MSV as a platform where pairwise entanglement, collective Gaussian steering, and collective non-Gaussianity separate cleanly. The proposed collective-mode subtraction protocol gives a direct route to heralded remote preparation of Wigner-negative states. Future work should address asymmetric coupling, energy-normalized comparisons with chain-type C3MSV resources, correlated losses in realistic FWM media, and a full experimental error model for collective-mode subtraction and tomography, including detector inefficiency and dark counts relevant to remote Wigner-negativity experiments~\cite{Liu2022}.

\begin{acknowledgments}
This work is supported by the National Natural Science Foundation of China (Grant No.~12564049), the Jiangxi Provincial Natural Science Foundation (Grant No.~20242BAB26009), the Jiangxi Provincial Key Laboratory of Advanced Electronic Materials and Devices (Grant No.~2024SSY03011), and the Jiangxi Civil-Military Integration Research Institute (Grant No.~2024JXRH0Y07).
\end{acknowledgments}

\appendix

\section{Bogoliubov transformation matrices (detailed)}
\label{app:A}

We present a pedagogical derivation of the Bogoliubov matrices $u$ and $v$. The squeezing operator is $S = \exp[\frac{1}{2}\bm{a}^\dagger M \bm{a}^{\dagger T} - \frac{1}{2}\bm{a}^T M^*\bm{a}]$, where $M$ is the $3\times3$ symmetric coupling matrix with $M_{ii}=0$ and $M_{i\neq j}=\xi=re^{i\theta}$.

\textit{Step 1: Fourier diagonalization.} $M$ is diagonalized by the orthogonal matrix
\begin{equation}
Q = \begin{pmatrix}
\frac{1}{\sqrt{3}} & \frac{1}{\sqrt{2}} & \frac{1}{\sqrt{6}}\\[2pt]
\frac{1}{\sqrt{3}} & -\frac{1}{\sqrt{2}} & \frac{1}{\sqrt{6}}\\[2pt]
\frac{1}{\sqrt{3}} & 0 & -\frac{2}{\sqrt{6}}
\end{pmatrix},
\end{equation}
which satisfies $Q^TQ=I_3$. The first column $\frac{1}{\sqrt{3}}(1,1,1)^T$ is the symmetric eigenvector; the other two columns span the antisymmetric subspace. One verifies $M = Q\operatorname{diag}(2\xi,-\xi,-\xi)Q^T$ by direct matrix multiplication. Physically, the eigenvalue $2\xi$ corresponds to the symmetric (center-of-mass) mode being squeezed with strength $2r$, while the doubly degenerate eigenvalue $-\xi$ corresponds to the two antisymmetric modes squeezed with strength $-r$.

\textit{Step 2: Polar decomposition.} Writing $M = H e^{iF}$ with $H$ Hermitian and $e^{iF}$ unitary: $H = Q\operatorname{diag}(2r,r,r)Q^T$, $e^{iF} = Q\operatorname{diag}(e^{i\theta},-e^{i\theta},-e^{i\theta})Q^T$. The negative signs in the second and third entries of $e^{iF}$ arise from the negative eigenvalues of $M$ and ensure $e^{iF}$ is unitary.

\textit{Step 3: Computing $u$ and $v$.} Following the similarity transformation theory of Fan and Zou~\cite{Fan1999}, $u=\cosh H$ and $v = e^{iF}\sinh H$ (the sign convention is chosen to match the main-text Bogoliubov transformation, which produces the positive $x$--$x$ correlations characteristic of the squeezing Hamiltonian~\eqref{eq:H}). Using the spectral decomposition:
\begin{align}
u &= Q\operatorname{diag}(\cosh2r,\cosh r,\cosh r)Q^T, \\
v &= Q\operatorname{diag}(e^{i\theta},-e^{i\theta},-e^{i\theta})\operatorname{diag}(\sinh2r,\sinh r,\sinh r)Q^T \nonumber\\
  &= Q\operatorname{diag}(e^{i\theta}\sinh2r,\,-e^{i\theta}\sinh r,\,-e^{i\theta}\sinh r)Q^T.
\end{align}
The resulting $v$ matrix matches Eqs.~\eqref{eq:v}--\eqref{eq:uv1} of the main text.

\textit{Step 4: Matrix elements.} Writing $u_{ij} = \sum_{k=1}^3 Q_{ik}\lambda_k^{(u)} Q_{jk}$ and similarly for $v$, where $\lambda^{(u)} = (\cosh2r,\cosh r,\cosh r)$ and $\lambda^{(v)} = (e^{i\theta}\sinh2r,\,-e^{i\theta}\sinh r,\,-e^{i\theta}\sinh r)$. The diagonal element:
\begin{align}
u_{11} &= \frac{1}{3}\cosh2r + \frac{1}{2}\cosh r + \frac{1}{6}\cosh r \nonumber\\
&= \frac{\cosh2r+2\cosh r}{3},
\end{align}
and the off-diagonal element:
\begin{align}
u_{12} &= \frac{1}{3}\cosh2r - \frac{1}{2}\cosh r + \frac{1}{6}\cosh r \nonumber\\
&= \frac{\cosh2r-\cosh r}{3}.
\end{align}
By cyclic symmetry, all diagonal elements equal $u_0/3$ and all off-diagonal elements equal $u_1/3$, with $u_0=\cosh2r+2\cosh r$ and $u_1=\cosh2r-\cosh r$. Similarly, $v_{ij} = e^{i\theta}v_0/3$ (diagonal) and $e^{i\theta}v_1/3$ (off-diagonal) with $v_0=\sinh2r-2\sinh r$ and $v_1=\sinh2r+\sinh r$, yielding Eqs.~\eqref{eq:u}--\eqref{eq:v} of the main text.

\textit{Step 5: Verification.} One checks $uu^\dagger-vv^\dagger=I_3$ and $uv^T=vu^T$ using $\cosh^2x-\sinh^2x=1$ and the orthogonality of $Q$. For example, the first condition in the eigenbasis reads $\cosh^2\lambda_k - \sinh^2\lambda_k = 1$ for each eigenvalue $\lambda_k$, which is identically satisfied.

\section{Normal ordering of the three-mode squeezing operator (detailed)}
\label{app:B}

We provide a pedagogical derivation of the normally ordered form of $S$ via the IWOP technique~\cite{Fan1987,Fan1999}. The goal is to express $S$ as $S = c\; {:}\!\exp(\text{quadratic in }a_i^\dagger,a_i)\!{:}$, where ${:}\cdot{:}$ denotes normal ordering (all $a_i^\dagger$ to the left of all $a_i$). This form is essential for computing Fock-space expansions and transition amplitudes.

\textit{Step 1: Quadratic form.} Define $A=(a_1,a_2,a_3)^T$, $A^\dagger=(a_1^\dagger,a_2^\dagger,a_3^\dagger)^T$, and the $6$-component vector $B=(A^\dagger, A)$. The squeezing operator is
\begin{equation}
S = \exp\Bigl[\frac{1}{2}B\,\Gamma\,\tilde{B}\Bigr],\qquad
\Gamma = \begin{pmatrix} R & 0 \\ 0 & D \end{pmatrix},
\end{equation}
where $R=\xi\mathbf{J}$, $D=-\xi^*\mathbf{J}$, $\mathbf{J}=\mathbf{1}\mathbf{1}^T-I_3$, $\mathbf{1}=(1,1,1)^T$, and $\tilde{B}=(-A, A^\dagger)^T$. The minus sign in $\tilde{B}$ is the standard symplectic convention used in Fan's normal-ordering formula~\cite{Fan1999}. The matrix $\mathbf{J}$ has eigenvalues $2$ (non-degenerate) and $-1$ (doubly degenerate), with eigenvectors $\frac{1}{\sqrt{3}}(1,1,1)^T$ and the two vectors in the orthogonal complement.

\textit{Step 2: Symplectic matrix.} Fan's key result~\cite{Fan1999} is that $S$ satisfies $S B S^{-1} = B M$, where $M = \exp(\Gamma\Pi)$ with the symplectic unit $\Pi = \begin{pmatrix}0 & -I_3 \\ I_3 & 0\end{pmatrix}$. Computing $\Gamma\Pi$:
\begin{equation}
\Gamma\Pi = \begin{pmatrix} 0 & -\xi\mathbf{J} \\ -\xi^*\mathbf{J} & 0 \end{pmatrix}.
\end{equation}
Diagonalizing $\mathbf{J}$ using the projectors $P_+ = \frac{1}{3}\mathbf{1}\mathbf{1}^T$ (onto the symmetric subspace) and $P_- = I_3-P_+$ (onto the antisymmetric complement), we have $\mathbf{J} = 2P_+ - P_-$.

\textit{Step 3: Exponentiation.} The block structure of $\Gamma\Pi$ decouples in the eigenbasis of $\mathbf{J}$. In the symmetric subspace, the $2\times2$ block $\begin{pmatrix}0 & -2\xi \\ -2\xi^* & 0\end{pmatrix}$ exponentiates to:
\begin{equation}
\begin{pmatrix} \cosh2r & -e^{i\theta}\sinh2r \\ -e^{-i\theta}\sinh2r & \cosh2r \end{pmatrix},
\end{equation}
yielding the symmetric-mode contributions $P = \cosh2r\,P_+$, $L = -e^{i\theta}\sinh2r\,P_+$. In the antisymmetric subspace (doubly degenerate), the block $\begin{pmatrix}0 & \xi \\ \xi^* & 0\end{pmatrix}$ exponentiates to give $P = \cosh r\,P_-$, $L = e^{i\theta}\sinh r\,P_-$. Assembling:
\begin{align}
P &= \cosh2r\,P_+ + \cosh r\,P_- \nonumber\\
  &= \cosh r\,I_3
   + \frac{\cosh2r-\cosh r}{3}\mathbf{1}\mathbf{1}^T, \\
L &= e^{i\theta}[-\sinh2r\,P_+ + \sinh r\,P_-], \qquad N = L^*.
\end{align}

\textit{Step 4: Normal ordering formula.} Equation~(14) of Ref.~\cite{Fan1999} gives the normally ordered form for any similarity operator with symplectic matrix $\begin{pmatrix}Q & L \\ N & P\end{pmatrix}$:
\begin{multline}
S = \frac{1}{\sqrt{\det P}}\,{:}\!\exp\Bigl[-\tfrac{1}{2}A^\dagger(LP^{-1})\tilde{A}^\dagger \\
+ A^\dagger(P^{-1}-I)\tilde{A} + \tfrac{1}{2}A(P^{-1}N)\tilde{A}\Bigr]\!{:}.
\end{multline}
The prefactor $1/\sqrt{\det P}$ gives the correct vacuum amplitude $\langle 0|S|0\rangle = 1/\sqrt{\det P}$; it reduces to unity only for $r=0$. Computing the needed quantities:
\begin{itemize}
\item $\det P = \cosh2r\cosh^2 r$,
\item $P^{-1} = \operatorname{sech}2r\,P_+ + \operatorname{sech}r\,P_-
      = \operatorname{sech}r\,I_3
      + [(\operatorname{sech}2r-\operatorname{sech}r)/3]\mathbf{1}\mathbf{1}^T$,
\item $LP^{-1} = e^{i\theta}[-\tanh2r\,P_+ + \tanh r\,P_-]$,
\item $P^{-1}-I = (\operatorname{sech}2r-1)P_+ + (\operatorname{sech}r-1)P_-$.
\end{itemize}

\textit{Step 5: Expansion in the original basis.} Substituting $P_+ = \frac{1}{3}\mathbf{1}\mathbf{1}^T$ and $P_- = I_3 - \frac{1}{3}\mathbf{1}\mathbf{1}^T$, and expanding the quadratic forms in terms of $a_i$ and $a_i^\dagger$, yields:
\begin{align}
-\frac{1}{2}A^\dagger(LP^{-1})\tilde{A}^\dagger &= \mathcal{P},\\
A^\dagger(P^{-1}-I)\tilde{A} &= \mathcal{N},\\
\frac{1}{2}A(P^{-1}N)\tilde{A} &= -\mathcal{P}^\dagger,
\end{align}
with $\mathcal{P}$, $\mathcal{N}$, and $\mathcal{P}^\dagger$ as given in the main text. The minus sign in front of $\mathcal{P}^\dagger$ reflects $N=L^*$ and the symmetry of $P^{-1}$. The final result is Eq.~\eqref{eq:normal_S}. For $r\to0$, $\mathcal{P},\mathcal{N},\mathcal{P}^\dagger\to0$ and the prefactor $\to1$, recovering $S\to I$.

\section{Covariance matrix calculation (detailed)}
\label{app:C}

We provide a complete, self-contained derivation of the covariance matrix from the Bogoliubov transformation. For the FC-C3MSV $|\psi\rangle=S|0\rangle$, all first moments vanish, so the CM is fully determined by the second moments.

\textit{Step 1: Second moments in terms of $u$ and $v$.} The Bogoliubov transformation $S^\dagger a_i S = \sum_j(u_{ij}a_j + v_{ij}a_j^\dagger)$ gives the second moments:
\begin{align}
\langle a_i^\dagger a_j\rangle &= \sum_{k,l} v_{ik}^* v_{jl} \langle 0|a_k a_l^\dagger|0\rangle = (v^\dagger v)_{ji}, \\
\langle a_i a_j\rangle &= \sum_{k,l} u_{ik} v_{jl} \langle 0|a_k a_l^\dagger|0\rangle = (uv^T)_{ij}, \\
\langle a_i a_j^\dagger\rangle &= \delta_{ij} + \langle a_j^\dagger a_i\rangle = \delta_{ij} + (v^\dagger v)_{ij},
\end{align}
using $\langle 0|a_k a_l^\dagger|0\rangle=\delta_{kl}$ (terms with $a_k^\dagger|0\rangle$ vanish).

\textit{Step 2: Computing $(v^\dagger v)_{ij}$.} For $\theta=0$, $v$ is real. The diagonal:
\begin{align}
(v^\dagger v)_{ii} &= \frac{v_0^2 + 2v_1^2}{9} \nonumber\\
&= \frac{(\sinh2r-2\sinh r)^2 + 2(\sinh2r+\sinh r)^2}{9} \nonumber\\
&= 2\sinh^2 r + \frac{4}{3}\sinh^4 r.
\end{align}
The off-diagonal element ($i\neq j$):
\begin{align}
(v^\dagger v)_{ij} = \sum_{k=1}^3 v_{ki}v_{kj} = \frac{2v_0v_1 + v_1^2}{9}.
\end{align}

\textit{Step 3: Computing $(uv^T)_{ij}$.} For $\theta=0$, $u$ and $v$ are both real. The diagonal:
\begin{align}
(uv^T)_{ii} = \sum_{k=1}^3 u_{ik}v_{ik} = \frac{u_0v_0 + 2u_1v_1}{9}.
\end{align}
Substituting the expressions for $u_0,u_1,v_0,v_1$ and simplifying using $\cosh2r=\cosh^2r+\sinh^2r$, $\sinh2r=2\sinh r\cosh r$:
\begin{align}
(uv^T)_{ii} = \frac{4}{3}\sinh^3 r\cosh r.
\end{align}
The off-diagonal ($i\neq j$):
\begin{align}
(uv^T)_{ij} = \frac{u_0v_1 + u_1v_0 + u_1v_1}{9}.
\end{align}

\textit{Step 4: Quadrature variances.} Expanding $\hat{x}_i^2 = \frac{1}{2}(\hat{a}_i+\hat{a}_i^\dagger)^2$ and taking expectation values, the cross terms simplify for $\theta=0$ (real $u$, $v$) giving $\langle\hat{a}_i^2\rangle=\langle\hat{a}_i^{\dagger2}\rangle=(uv^T)_{ii}$. Hence:
\begin{equation}
\langle \hat{x}_i^2\rangle = \tfrac{1}{2}\bigl(1 + 2(v^\dagger v)_{ii} + 2(uv^T)_{ii}\bigr).
\end{equation}
Similarly, $\langle \hat{p}_i^2\rangle = \frac{1}{2}(1 + 2(v^\dagger v)_{ii} - 2(uv^T)_{ii})$.

\textit{Step 5: Cross-correlations.} For $i\neq j$ (e.g., $i=1,j=2$) at $\theta=0$:
\begin{equation}
\langle \hat{x}_i\hat{x}_j\rangle = (v^\dagger v)_{ij} + (uv^T)_{ij},
\end{equation}
using the expansion $\hat{x}_i\hat{x}_j = \frac{1}{2}(\hat{a}_i+\hat{a}_i^\dagger)(\hat{a}_j+\hat{a}_j^\dagger)$ and noting that cross terms $\langle\hat{a}_i\hat{a}_j^\dagger\rangle$ combine with $(v^\dagger v)_{ij}$. The $x$-$p$ cross-terms vanish since all quantities are real at $\theta=0$.

\textit{Step 6: Assembling $A$ and $B$.} The single-mode block $A$ has diagonal entries $A_{11}=\langle\hat{x}_1^2\rangle$, $A_{22}=\langle\hat{p}_1^2\rangle$ (off-diagonals vanish at $\theta=0$). The inter-mode block $B$ has $B_{11}=\langle\hat{x}_1\hat{x}_2\rangle$, $B_{22}=\langle\hat{p}_1\hat{p}_2\rangle$. Converting to exponential form via $\cosh x=(e^x+e^{-x})/2$, $\sinh x=(e^x-e^{-x})/2$ yields Eqs.~\eqref{eq:A}--\eqref{eq:B}.

\textit{Step 7: Purity check.} The Fourier-block symplectic eigenvalues are:
\begin{equation}
\nu_0 = \sqrt{\det(A+2B)} = \tfrac{1}{2},\qquad
\nu_1 = \sqrt{\det(A-B)} = \tfrac{1}{2},
\end{equation}
confirming $|\psi\rangle=S|0\rangle$ is a pure Gaussian state for all $r\ge0$. For reference, the correspondence between the second-moment notation used above and the CM-block entries of Eqs.~\eqref{eq:A}--\eqref{eq:B} is: $a_x=A_{11}=1/2+(v^\dagger v)_{11}+(uv^T)_{11}$, $b_x=B_{11}=(v^\dagger v)_{12}+(uv^T)_{12}$, with analogous expressions for the $p$-quadrature entries.

\section{Normalized Wigner function after photon subtraction}
\label{app:D}

We derive the conditional Wigner function used in Sec.~\ref{sec:wn}. Consider the two-mode zero-mean Gaussian Wigner function
\begin{equation}
W_{am}(\bm r)=\frac{1}{4\pi^2\sqrt{\det V_{am}}}
\exp\!\left[-\frac{1}{2}\bm r^T V_{am}^{-1}\bm r\right],
\end{equation}
where $\bm r=(x_a,p_a,x_m,p_m)^T$ and $V_{am}$ is given by Eq.~\eqref{eq:Vam}. The marginal of mode $a$ is
\begin{equation}
W_a^G(x_a,p_a)=\frac{1}{2\pi\sqrt{a_xa_p}}
\exp\!\left[-\frac{x_a^2}{2a_x}-\frac{p_a^2}{2a_p}\right].
\end{equation}
The ideal photon-subtracted conditional state is
\begin{equation}
\tilde{\rho}_a=\operatorname{Tr}_m(\hat a_m\rho_{am}\hat a_m^\dagger)
=\operatorname{Tr}_m(\hat n_m\rho_{am}).
\end{equation}
In the $[\hat x,\hat p]=i$ convention, the Weyl symbol of $\hat n_m$ is
\begin{equation}
n_m(x_m,p_m)=\frac{x_m^2+p_m^2-1}{2}.
\end{equation}
Therefore
\begin{equation}
\tilde W_a(x_a,p_a)=\int dx_m\,dp_m\,
\frac{x_m^2+p_m^2-1}{2}W_{am}(\bm r).
\end{equation}
Conditioned on fixed $(x_a,p_a)$, the measured-mode Gaussian has means
\begin{equation}
\mu_x=\frac{c_x}{a_x}x_a,\qquad
\mu_p=\frac{c_p}{a_p}p_a,
\end{equation}
and conditional variances
\begin{equation}
d_{x|a}=d_x-\frac{c_x^2}{a_x},\qquad
d_{p|a}=d_p-\frac{c_p^2}{a_p}.
\end{equation}
Thus
\begin{align}
\tilde W_a(x_a,p_a)
&=W_a^G(x_a,p_a)
\frac{d_{x|a}+\mu_x^2+d_{p|a}+\mu_p^2-1}{2}\\
&=W_a^G(x_a,p_a)
\left(C_m+K_xx_a^2+K_pp_a^2\right),
\label{eq:Wa_unnormalized_app}
\end{align}
with $C_m,K_x,K_p$ as defined in Eq.~\eqref{eq:CmPm}. Integrating Eq.~\eqref{eq:Wa_unnormalized_app} over the target phase space gives
\begin{align}
P_m&=C_m+K_xa_x+K_pa_p\\
&=\frac{d_x+d_p-1}{2}=\langle \hat n_m\rangle.
\end{align}
The normalized Wigner function is therefore Eq.~\eqref{eq:Wa_general}.

For a single physical measured mode $m=b$, $d_x=a_x$, $d_p=a_p$, $c_x=b_x$, and $c_p=b_p$. In the lossless case,
\begin{align}
d_{x|a}^{(b)}&=a_x-\frac{b_x^2}{a_x}
=\frac{e^{-2r}(2e^{6r}+1)}{2(e^{6r}+2)},\\
d_{p|a}^{(b)}&=a_p-\frac{b_p^2}{a_p}
=\frac{e^{2r}(e^{6r}+2)}{2(2e^{6r}+1)}.
\end{align}
Their product is $1/4$, so $d_{x|a}^{(b)}+d_{p|a}^{(b)}\ge1$ and $C_b\ge0$. Hence photon subtraction on a single physical mode cannot produce Wigner negativity in another physical mode of the symmetric FC-C3MSV.

For the collective mode $d=(b+c)/\sqrt{2}$, the entries are those in Eq.~\eqref{eq:collective_entries}. At $r=0.5$ and $\eta=1$,
\begin{equation}
d_{x|a}^{(d)}=0.5018,\quad d_{p|a}^{(d)}=0.0990,
\quad C_d=-0.1996,
\end{equation}
which gives a negative Wigner value at the origin. Under full symmetric loss, solving $C_d(\eta)=0$ yields $\eta_c=0.6671605\ldots$.

Finally, we give the negative-volume formula used in Table~\ref{tab:wn}. Let
\begin{equation}
\begin{aligned}
x_a&=\sqrt{a_x}\rho\cos\phi,\\
p_a&=\sqrt{a_p}\rho\sin\phi,
\end{aligned}
\end{equation}
and define
\begin{equation}
L(\phi)=K_xa_x\cos^2\phi+K_pa_p\sin^2\phi.
\end{equation}
If $C_m\ge0$, then $W_a^{(m)}\ge0$ everywhere and $\mathcal{N}=0$. If $C_m<0$, the negative region is $0\le\rho^2<-C_m/L(\phi)$, and
\begin{equation}
\mathcal{N}=-\frac{1}{\pi P_m}\int_0^{2\pi}
\left[C_m+2L(\phi)\left(1-e^{C_m/[2L(\phi)]}\right)\right]d\phi.
\label{eq:neg_volume_general}
\end{equation}
This expression is finite at the threshold, where it tends continuously to zero.

\end{document}